\documentclass[a4paper,11pt]{article}
\usepackage[T1,T2A]{fontenc}
\usepackage[utf8]{inputenc}
\usepackage{amsmath,amsthm,amssymb}
\usepackage{graphics}
\usepackage{graphicx}
\begin{document}
\title{Spin relaxation in a magnetic field in Al}
\author{Yu.N. Chiang (Tszyan), M.O. Dzyuba\\
\emph{B. I. Verkin Institute for Low Temperature Physics and Engineering,}\\
\emph{National Academy of Sciences of Ukraine.}\\
\emph{Nauky Ave. 47, Kharkov 61103, Ukraine}\\
\emph{E-mail:chiang@ilt.kharkov.ua}}
\date{}
\maketitle
\begin{abstract}
The direct electrical method was used to study the behavior of the spin Hall effect in a magnetic field perpendicular to the injection current in aluminum samples with resistivities that differed by two orders of magnitude at a temperature of 4.2 K. The parabolic behavior of the spin-Hall voltage curves with maxima at the same value of the magnetic field was discovered. It was proposed an explanation which is based on the competition of two universal mechanisms associated with the accumulation of spins - the dependences of spin magnetization and spin relaxation length on the magnetic field strength.
\end{abstract}

The problem of relaxation of nonequilibrium spin polarization occupies an important place in the theory and practice of spintronics. In particular, this applies to non-magnetic metals, in which spin-orbit interaction (SOI) plays an important role in the spin splitting of the conduction band. The latter can significantly change the electron dispersion law, leading to the appearance of an effective magnetic field, rigidly linked to the wave vector of the charge carrier and acting on the electron spin, which causes the dependence of the electron velocity direction on the spin direction and the appearance of the spin-Hall effect (SHE) [1]. Since the dynamics of spins of opposite orientations are opposite in sign and are equally probable, the occurrence of a transverse nonequilibrium spin-charge imbalance, which causes SHE in a non-magnetic sample, is, generally speaking, impossible unless a obviously non-degenerate and spin-nonequilibrium current is injected (for example, from a ferromagnet [2]), or resort to indirect methods of its creation [3, 4]. Indeed, if the Hall cross is defined in the plane ($x,y$) of the sample, and the current is passed in the direction $\pm~ x$, then the spin-charge voltage in the direction \emph{y} is defined as
\begin{equation}\label{1}
    V_{\mathrm{SHE}} \sim ~[\mathbf{\hat{\sigma}}\times\mathbb{\nabla} (\mu_{\uparrow}+\mu_{\downarrow})]_{y},
\end{equation}
where $\mu_{\uparrow}~ and ~\mu_{\downarrow}$ are the chemical potentials of charge carriers in the \emph{y} direction with the corresponding spin orientation along \emph{z},~ $\mathbf{\hat{\sigma }}$ are Pauli matrices. For sample width $L_{y} < \lambda_{sf}$ (spin relaxation length) $\mu_{\uparrow}= - \mu_{\downarrow}$ and $V_{SH}\equiv 0$ for any sample shape . However, we showed [5] that for $L_{y} > \lambda_{sf}$ and sample asymmetry in the direction \emph{y}$~|\mu_{\uparrow}|\neq |\mu_{\downarrow }|$, so that $V_{SH}\neq 0$, which suggests a direct method for electrical measurements of the spin Hall effect in paramagnetic metals under conditions of injection of a degenerate electron flow. Figure 1 explains these conditions.

\begin{figure}    .
  \centering
  \includegraphics[width=10cm]{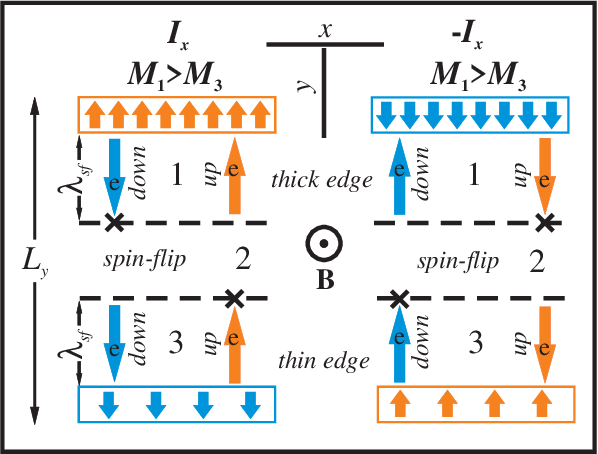}
  \caption{The process of spin accumulation in an asymmetric paramagnetic sample with the same sign of carriers, cross section $A_{xz}$ and width $L_{y}$ exceeding the spin relaxation length $\lambda_{sf}$. Crosses indicate spin fluxes that lose polarization.}\label{1}
  \end{figure}

Here we study SHE in an external magnetic field, which has a specific influence on the behavior of this effect, which allows us to further identify the effect as a spin one. The validity of considering spin effects associated with spin-orbit interaction, in particular, in a bulk metal, is, as is known, due to the removal of the center of spatial inversion by an external field \textbf{E}. In this case, the magnetic field \textbf{B} has a versatile effect on the dynamics of charges and their spins. On the one hand, it leads to spin-charge accumulation of carriers at the edges of the sample with non-parallel orientation of \textbf{E} and \textbf{B} relative to each other [1], but, on the other hand, it destroys the spin polarization, simultaneously changing the initial spin magnetization, arising as a result of the accumulation of spins. Under these conditions, as discussed earlier [5, 6], the elicitation of the contribution directly related to spin effects requires a number of restrictions, in particular regarding to the choice of field interval and experiment geometry. Of course this is due to the large contribution of the usual Hall effect, which, fortunately, is a strictly odd effect with respect to the magnetic field and injection current in the geometry $\mathbf{B}\bot\mathbf{E}$, while SHE is even in the electric current (Fig. 1) and in the magnetic field for low values.

\begin{figure}
  \centering
  \includegraphics[width=15cm]{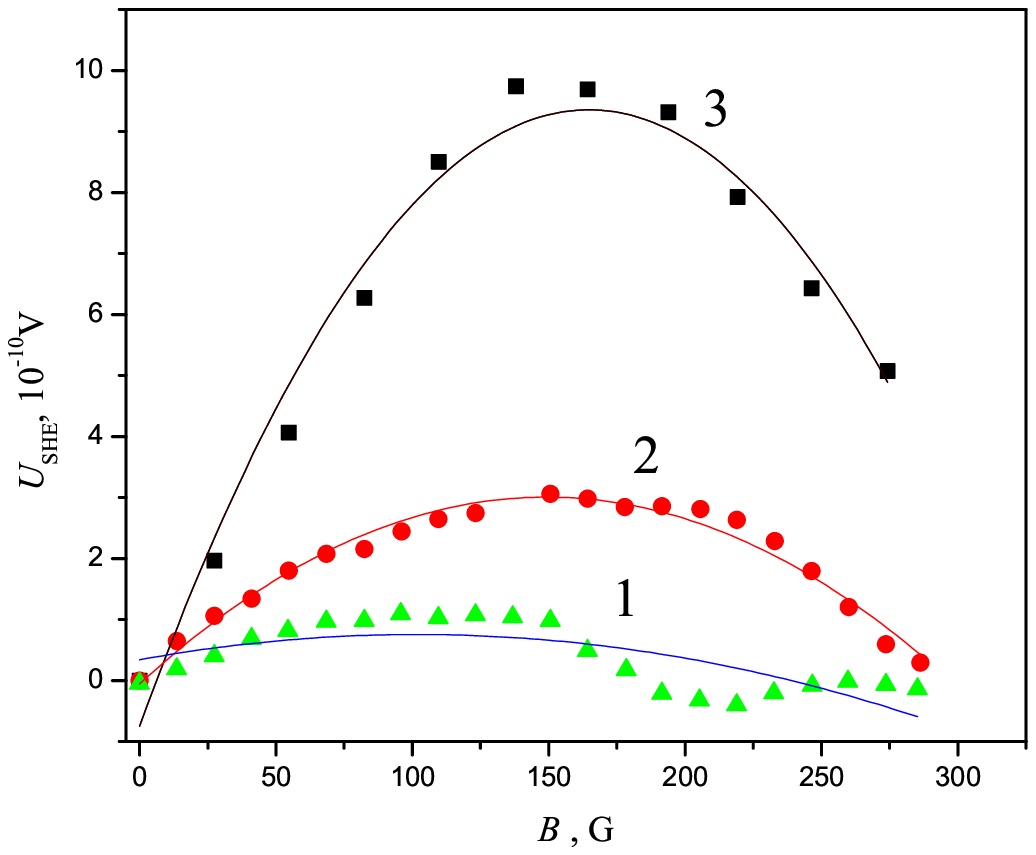}\vspace{-3cm}
  \caption{Spin-Hall effect in 3 asymmetric aluminum samples of different purities in a transverse magnetic field. Symbols are experiment, solid curves are fitted using expression (6).
  }\label{2}
\end{figure}

As is known, the interaction of spin with a magnetic field is described by the Hamiltonian
\begin{equation}\label{2}
    \mathcal{H}_{B}= g\mu_{B}(\hat{\sigma}\cdot B)= \frac{\hbar}{2}(\hat{\sigma}\cdot B),
\end{equation}
where \emph{g} is the g-factor, $\mu_{B}$ is the Bohr magneton. From this follows the classical interpretation of the precession equation for the averaged spin component $\emph{\textbf{S}}$:
\begin{equation}\label{3}
    \frac{\emph{d\textbf{S}}}{\mathrm{d}t}=\mathbf{\Omega}\times \emph{\textbf{S}},
\end{equation}
with angular precession frequency $\mathrm{\Omega}=|\mathbf{\Omega}|=g\mu_{B}B/\hbar$. In this interpretation, the precession period $T=2\pi\Omega^{-1}$ corresponds to the spin relaxation time $\tau_{s}$, i.e. the time at which the spin direction is lost. Indeed, if electrons are scattered elastically on defects with a momentum relaxation time $\tau_{c}$, then the spin deflection angle at the electron mean free path $\ell_{c}=\tau_{c}v_{\mathrm{F}}$ between successive collisions will be $\delta\vartheta=\Omega\tau_{c}$, which follows ([7]):
\begin{equation}\label{4}
    \frac{1}{\tau_{s}} \simeq\Omega^{2}\tau_{c}
\end{equation}
and at $\Omega\tau_{c}\sim 1$ we have $\tau_c\sim \tau_s\sim T$, i.e. the precession period is the time of spin depolarization, and expression (4) indicates the dependence of the spin relaxation length on the magnetic field value and momentum relaxation time $\tau_{c}$. Figure 2 shows the field dependences in a perpendicular magnetic field of the spin contribution to the transverse voltage $U_{y}^{s}$ for three asymmetric aluminum samples of different purities with $\tau_{c} (4.2K): 5.8\cdot 10^{-11}s; 1.8\cdot 10^{-11}s; 1.8\cdot 10^ {-13}s$ (in the order of numbering the curves in the figure) after excluding the standard Hall voltage. The correctness of the exclusion is justified by the small field interval: $\omega_{c}\tau_{c}\leq 0.1$, where $\omega_{c}$ is the cyclotron frequency. We see that, regardless of the purity of the samples, the curves have a peak at almost the same value of the magnetic field ($\approx$ 160 G), indicating that its position is determined only by the spin precession frequency at a fixed \emph{g}-factor (equal to 2 for free electrons), and the presence of two competing independent spin mechanisms associated with the magnetic field. We consider that such mechanisms may be an increase in spin magnetization \emph{\textbf{M}}(\textbf{B}), on the one hand, and a decrease in the spin precession period, entailing a decrease in the spin relaxation length, on the other hand, with increasing field :
\begin{equation}\label{5}
    \mathcal{E}_{y}^{z}=\frac{U_{y0}}{\lambda_{sf0}} - \frac{U_{y0}}{\lambda_{sf}} + \frac{U_{\emph{\textbf{M}}}(\Delta \emph{\textbf{M})}}{L_{y}}
\end{equation}
where $\lambda_{sf}\sim \tau_{s}\sim (\Omega^{2}\tau_{c})^{-1}$, and $\Delta\emph{\textbf{M}}= (\emph{\textbf{M}}_{1}-\emph{\textbf{M}}_{3})\sim \mathbf{B}$ (for designations see fig. 1), so the curves in Fig. 2 can be described by the expression
\begin{equation}\label{6}
    U_{\mathrm{SHE}} = U_{y0} - \alpha(\tau_{c}) B^{2} + \beta_{\Delta\emph{\textbf{M}}} B,
\end{equation}
as shown by the solid fit curves in Fig. 2.

There is a difference between the experimental behavior of curve 1 (points) and the dependence described by the second term of this expression in the region of decreasing spin relaxation length. In addition to the assumptions we proposed in [5], this behavior can also be associated with a noticeable value of the effective field $\omega_{c}\tau_{c}\sim 0.1$ (two orders of magnitude greater than that of curve 3). As a result, the spin relaxation length could decrease with increasing field more effectively:
\begin{equation}\label{7}
    \lambda_{sf}\sim [\Omega^{2}\tau_{c}(1+\omega_{c}^{2}\tau_{c}^{2})]^{-1}.
\end{equation}

It is interesting to compare the position of the considered peak on the $U_{\mathrm{SHE}}$ curves for aluminum with the position of a similar peak for platinum. In the last one observing visually on oscillograms the manifestation of the spin-Hall effect on alternating current $\mathcal{I}_{x}= \mathcal{I}_{x0}\sin \omega t$ [8] in the form of the appearance of a double frequency signal relative to $\mathcal{I}_{x}$, we observed the maximum effect at $B_{max} = 313 G$. Let us remind that the reason for the frequency doubling is the simultaneous presence of two transverse contributions - spin and ordinary Lorentzian, independently depending on $\mathcal{I}_{x}: U_{y}= U_{\rm SHE}(t, B) [1 \pm \alpha_{m}(B)\sin\omega t] \sim \sin ^{2} \omega t\sim \cos 2\omega t$ ($\alpha_{m}(B) = \frac{U_{\rm CHE (0)}(B)}{U_{\rm SHE (0)}(B)}$) is modulation depth, proportional to the second term in expression (6)). Thus, we can assume that the SO coupling coefficient is no less than 2 times higher in Pt.

In conclusion,the behavior of the spin-Hall voltage in an asymmetric aluminum samples of various purities depending on the magnetic field is considered. It was found that it is parabolic in nature for all samples. This behavior is typical for ensembles of accumulated spins with competing dependences on the magnetic field - the spin magnetization and the spin relaxation length associated with the spin precession frequency. Since the positions of the maxima of the dependences for samples of different purities coincide (\textit{B} = 160 G), we can say that these mechanisms have a universal nature within the same material, and their contribution depends only on the magnitude of the magnetic field, at least in the field region $\omega_{c}\tau_{c}\leq 0.1$.

\end{document}